\documentclass[prd,twocolumn,amsmath,aps,floats,amssymb, floatfix,nofootinbib]{revtex4-1}

\usepackage{graphicx}
\usepackage{wasysym}
\usepackage{amsfonts}
\usepackage{subfigure}
\usepackage[table]{xcolor}
\newcommand{\Bm}[1]{\mbox{\boldmath $#1$}}  
\usepackage{ulem}                                                                                                
\definecolor{lightgray}{gray}{0.9}                                                                                                                
                                                                                                                 
\begin{document}

\title{Scalar runnings and a test of slow roll from CMB distortions}
\author{Brian A.\ Powell} 
\email[Email:]{brian.powell007@gmail.com}
\affiliation{Institute for Defense Analyses, Alexandria, VA 22311, USA}
\date{September 10, 2012}

\begin{abstract}
\noindent 
A future measurement of cosmic microwave $\mu$-distortions by an experiment with
the specifications of PIXIE will provide an equivalent 3$\sigma$
detection of the running of running of the spectral index of
scalar perturbations, $\beta = d \alpha/d \ln k$, if $\mu > 7.75\times
10^{-8}$, covering much of the PIXIE
sensitivity range. This corresponds to a
resolution limit of $\beta \gtrsim 0.015$ which is relatively large given any presumption of slow roll, a result
of the current tight constraints on $\alpha < 0$ on CMB scales.  We show
that a detection of $\beta$ at this level is in
conflict with slow roll conditions if the primordial signal can be distinguished from any
post-inflationary contamination. 
\end{abstract}

\maketitle

\section{Introduction}
Understanding the character of the primordial density perturbation has been a chief endeavor of modern precision cosmology.  Our knowledge of the shape and
statistics of the primordial power spectrum has grown substantially over the past
two decades, from the Cosmic Background Explorer's first glimpse of the cosmic microwave background
(CMB) anisotropies, to today's swarm of
ground-based and balloon-borne CMB and large scale structure observatories.  Our knowledge of the
primordial density perturbations derives almost exclusively from these data sources -- the
temperature and polarization anisotropies of the cosmic microwave background and galaxy
correlation measurements -- both of which probe large length scales.  These observations have
revealed a nearly scale invariant spectrum of Gaussian, adiabatic perturbations on
comoving scales $k \lesssim 0.1$ Mpc$^{-1}$, in agreement with the simplest models of inflation.   

But inflation might not have been so simple. Deviations from Gaussianity and adiabaticity are
possible, and the power spectrum might deviate from a power law \cite{Vazquez:2012ux,Lesgourgues:2007gp,Powell:2007gu}.  
A spectrum with a nontrivial scale dependence
might indicate a
complicated inflaton potential exhibiting non-slow roll behavior.  A correspondence exists between the scale dependence of the
power spectrum, $P(k)$, and the field-dependence of the inflaton potential, $V(\phi)$: specifically, the n$^{th}$-order term of the Taylor expansion of $V(\phi)$ in
$\phi$ is linear in the n$^{th}$-order term of the Taylor expansion of $P(k)$ in $\ln k$
\cite{Lidsey:1995np}. By resolving the scale dependence of $P(k)$, in particular the higher-order runnings of the
spectral index: $\alpha = d n_s/d\rm lnk$, $ \beta = d \alpha/d\rm lnk$, etc., the paradigm of single field,
slow roll inflation can be tested.

In order to observe these higher-order runnings, the spectrum must be measured across a wide range of length scales.
The Wilkinson Microwave Anisotropy Probe (WMAP) and Sloan's observations of baryon acoustic oscillations
combine to probe scales $10^{-4}$ Mpc$^{-1} \lesssim k \lesssim 0.1$ Mpc$^{-1}$; while these
data exhibit a preference for $\alpha < 0$, it is not statistically significant
\cite{Komatsu:2010fb}.  
The Lyman alpha forest, which probes large scale structure on scales $k\approx 1$
Mpc$^{-1}$ \cite{Viel:2004np,McDonald:2004xn,Seljak:2006bg},
is promising, but converting flux measurements to a density power spectrum is difficult with
our currently incomplete understanding of the properties of the intergalactic medium.
Recent ground-based observatories, like the South Pole Telescope\footnote{http://pole.uchicago.edu/}and the Atacama
Cosmology Telescope\footnote{http://www.physics.princeton.edu/act/}, have begun to place nearly decisive limits on the running of the
spectral index by probing the damping tail of CMB fluctuations:
$\alpha = -0.034 \pm 0.018$ \cite{Dunkley:2010ge} and $\alpha = -0.024 \pm 0.0112$ \cite{Keisler:2011aw}, and Planck will
do even better. 
Current data, however, can go no further: in order to resolve higher-order runnings, we need a longer lever arm.
By extending the observational window to smaller scales, $P(k)$ can be
recovered across a greater range of $\ln k$, and subtle higher-order terms in its Taylor
expansion will become relevant. 
One promising probe of small scale fluctuations is
the 21cm transition of neutral hydrogen, which will extend the observational window down to
scales $k \sim 100$ Mpc$^{-1}$ \cite{Loeb:2003ya}, and resolve $\alpha$ at the level of $10^{-4}$
\cite{Barger:2008ii,Adshead:2010mc}.
We can expect these results in the coming years.

But it might be possible to go even further: the thermal spectrum of the cosmic
microwave background radiation traces fluctuations on scales 50 Mpc$^{-1} \lesssim k \lesssim
10^4$ Mpc$^{-1}$, offering an unprecedented glimpse into the character of the
small-scale power spectrum.  This is a fundamentally different kind of measurement: rather than probing the fluctuations themselves, 
the focus is on the {\it dissipation} of these fluctuations into the baryon-photon plasma and the resulting deviation
of the thermal spectrum of the radiation field from that of a black body.
The proposed PIXIE \cite{Kogut:2011xw} mission will be sensitive to these
spectral distortions at a level useful for power spectrum reconstruction, and there has
been recent interest in its 
prospects for improving constraints on $\alpha$ \cite{Khatri:2011aj,Chluba:2012gq,Dent:2012ne,Chluba:2012we}.
In this paper we investigate the potential of such a measurement for resolving the running of
the running of the spectral index, $\beta = d\alpha/ d\ln k$.

\section{CMB distortions}
Fluctuations in the baryon-photon plasma below the Jeans scale oscillate as acoustic waves during radiation
domination.  The energy of these fluctuations is dissipated via diffusion damping, resulting
in a redistribution of this energy from the fluctuations into the baryon-photon plasma
\cite{Silk:1967kq,Sunyaev:1970er,Peebles:1970ag,Illarionov:1975,Coles,Daly,Hu:1992dc,Chluba:2011hw}.
At early times ($z \geq 2 \times 10^6$), efficient double Compton emission and
bremsstrahlung, and the
up-scattering of these photons by energetic electrons quickly thermalizes the baryon-photon plasma, preserving its
black body spectrum.
As the universe cools -- between redshifts $5\times 10^4 \lesssim z \lesssim 2\times10^6$ -- Compton scattering ensures statistical equilibrium of the radiation field, but the production of photons is insufficient to
maintain a black body: a frequency-dependent chemical potential, $\mu(\omega)$, develops and the spectrum shifts to that of
a Bose-Einstein distribution.  So, while small-scale perturbations are erased before
recombination, they leave behind a record of their existence: the energy lost by
the perturbations is injected into the CMB distorting the
spectrum away from a black body by an amount that depends on the shape and
amplitude of the power spectrum on the relevant scales.

The time evolution of the $\mu$-distortion amplitude obeys \cite{Hu:1992dc}
\begin{equation}
\frac{d\mu}{dt} \approx -\frac{\mu}{t_{DC}} + 1.4\frac{Q}{\rho_\gamma},
\end{equation}
where $t_{DC}$ is the double Compton thermalization time, and $Q/\rho_\gamma$ is
the fractional rate of energy dissipation, and double Compton
emission has been assumed the dominant source of photon production
\cite{Danese:1982,Hu:1992dc}.  This has the solution
\begin{equation}
\label{mu}
\mu \approx 1.4 \int_{z_\mu}^\infty
\frac{\exp\left[-\left(z/z_{DC}\right)^{5/2}\right]}{\rho_\gamma} \frac{d Q}{dz}dz
\end{equation}
where $z_\mu = 5\times 10^4$ and $z_{DC} = 1.98 \times 10^6$.
The energy density of the acoustic waves is $Q = 3\rho_\gamma c^2_s \langle \delta
_\gamma ({\bf x})^2\rangle/4$, where $c_s^2 \approx 1/3$ is the squared sound speed,
and the photon density perturbation, $\delta_\gamma(\bf x)$, has the ensemble
average,
\begin{equation}
\langle \delta_\gamma ({\bf x})^2 \rangle = \int \frac{d^3 k}{(2\pi)^3}P(k),
\end{equation}
where $P(k) = \Delta^2_\gamma P^i(k)$,
$\Delta^2_\gamma$ the transfer function, and $P^i(k)$ is the primordial spectrum
generated by inflation.  Since we are considering acoustic waves below the Jeans
scale during radiation domination, the transfer function is appropriate to modes that are well inside the horizon,
\begin{equation}
\Delta_\gamma \approx 3 \cos(k r_s)e^{-k^2/k_D^2},
\end{equation}
where $r_s$ is the sound horizon and the diffusion scale, $k_D$, varies with redshift during radiation domination as
\begin{equation}
\label{kD}
k_D = A_D^{-1/2} (1+z)^{3/2},
\end{equation}
where $A_D = 8\left(135 H_0 \sqrt{\Omega_r} n_{e0}\sigma_T\right)^{-1} = 5.92 \times 10^{10}$
Mpc$^{-1}$.  Here $n_{e0}$ is the free electron number density at $z=0$ and
$\sigma_T$ is the Thompson scattering cross section.  The portion of the power
spectrum, $P^i(k)$, that is primarily responsible for generating the $\mu$-distortion is set by the diffusion
scales at the beginning and end of the epoch, $5\times10^4 \lesssim z \lesssim 2\times
10^6$.  Equation (\ref{kD}) gives $k \in [50, 10^4]$ Mpc$^{-1}$.

With these results, the amount of $\mu$-distortion Eq. (\ref{mu}) can be
well-approximated by \cite{Chluba:2012we}
\begin{equation}
\label{int}
\mu \approx 2.2 \int_{k_{\rm min}}^\infty
P^i(k)\left[\exp\left(-\frac{k}{5400}\right) -
\exp\left(-\frac{k^2}{998.6}\right)\right] d \ln k,
\end{equation}
where $\cos^2(k r_s)$ has been replaced by $1/2$ -- its average value over one
oscillation.  In contrast to the anisotropy measurements for which CMB data on scale $\ell$ constrains the spectrum
at corresponding wavenumber, $k$,  
the $\mu$-distortion measurement comes in the form of a weighted integral
constraint on $P^i(k)$ over the $\mu$-distortion epoch.  

\section{Observing ${\Bm \beta}$}
The question of whether an observation detects a physical quantity, $\theta$, is a problem in hypothesis testing: are the data sufficient to
overturn the null hypothesis, in which $\theta =0$?  In Bayesian statistics, hypotheses are interpreted as models, and hypothesis testing proceeds via a 
comparison of the Bayesian evidence of the first model, $p({\bf d}|\mathcal{M}_0)$,  with that of the second, $p({\bf d}| \mathcal{M}_1)$,
\begin{equation}
\label{bayes_factor}
B_{01} = \frac{p({\bf d}|\mathcal{M}_0)}{p({\bf d}|\mathcal{M}_1)},
\end{equation}
where {\bf d} is the data and $\mathcal{M}$ the model.  The Bayesian evidence is the average of the likelihood function, $\mathcal{L}({\Bm
\theta}|{\bf d}, \mathcal{M})$, weighted by the prior probability of the parameters, ${\Bm \theta}$,
\begin{equation}
\label{evidence2}
p({\bf d}|\mathcal{M}) = \int \mathcal{L}({\Bm \theta}|{\bf d},\mathcal{M})\pi({\Bm \theta}|\mathcal{M})d{\Bm \theta}.
\end{equation}
Once the evidence of each model is computed, the Bayes factor Eq. (\ref{bayes_factor}) can be evaluated and inferences made based on the
Jeffrey's scale \citep{Jeffreys}: $\ln B_{01}< 1$ means that the data is insufficient to distinguish
the models, while $1<\ln B_{01} < 2.5$, $2.5<\ln B_{01} < 5$, and $\ln B_{01} \geq 5$ signify positive, moderate, and strong evidence, respectively, in favor of
$\mathcal{M}_0$. In what follows, we will quote results that satisfy the threshold for strong evidence in favor of
$\mathcal{M}_1$,
$|\ln B_{01}| \geq 5$, corresponding to a $0.993$ posterior probability for the favored model
\cite{Trotta:2005ar}.

We are interested in the prospects of detecting and constraining higher-order runnings of the scalar power spectrum, specifically, the running
of the running of the spectral index, $\beta = d\alpha/d{\rm ln}k$, with future observations with the
specifications of the PIXIE experiment \cite{Kogut:2011xw}, a proposed nulling polarimeter for the measurement of
primordial gravitational waves.  The model that we wish to study, $\mathcal{M}_1$, has a scalar power
spectrum given by 
\begin{equation}
\label{spec}
P^i(k) = A_s\left(\frac{k}{k_0}\right)^{n_s-1+ \frac{1}{2}\alpha \ln\frac{k}{k_0} + \frac{1}{6}\beta
\ln^2\frac{k}{k_0}}.
\end{equation}
  The base model, $\mathcal{M}_0$, has $\beta = 0$ and so is contained in $\mathcal{M}_1$ as a special case.  
This is a standard exercise in Bayesian model comparison: is the data sufficient to warrant the addition
of the parameter $\beta$?  Equivalently, the model comparison determines whether the likelihood of the
model fit to the data is sufficiently improved, beyond statistical coincidence, when the new parameter
is included.  Because the models $\mathcal{M}_0$ and $\mathcal{M}_1$ are nested, we make use of a
representation of the Bayes factor known as the {\it Savage-Dickey density ratio} \cite{Dickey:1971},    
\begin{equation}
\label{SDR}
B_{01} = \left.\frac{p(\beta|{\bf d},\mathcal{M}_1)}{\pi(\beta|\mathcal{M}_1)}\right|_{\beta = 0},
\end{equation}
where $p(\beta|{\bf d},\mathcal{M}_1)$ is the marginalized posterior distribution of $\beta$ and $\pi(\beta|\mathcal{M}_1)$ its prior probability under model $\mathcal{M}_1$, and both are evaluated at $\beta = 0$ in Eq. (\ref{SDR}).  The Savage-Dickey density ratio is a convenient route to the Bayes' factor that does not require an evaluation of the Bayesian evidence, which typically involves a high-dimensional integration that can be computationally demanding.  It is nonetheless exact for nested models with separable priors.

\begin{figure}
\includegraphics[width=\linewidth]{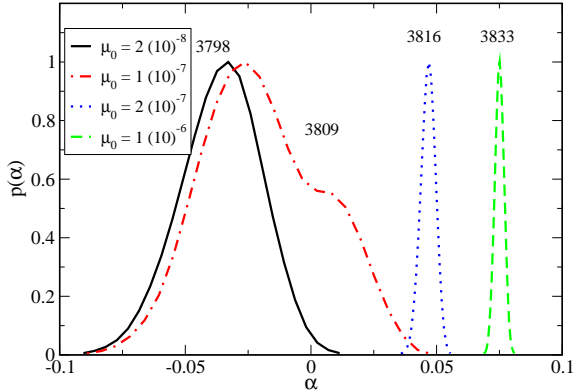}
\caption{Marginalized posteriors of $\alpha$ for different fiducial amplitudes of
$\mu$-distortion for model $\mathcal{M}_0$.  The
value of the best fit $\chi^2_{eff} = -2\ln \mathcal{L}$ is shown for each case.}
\end{figure}

\begin{figure*}[ht]
\includegraphics[width=5in]{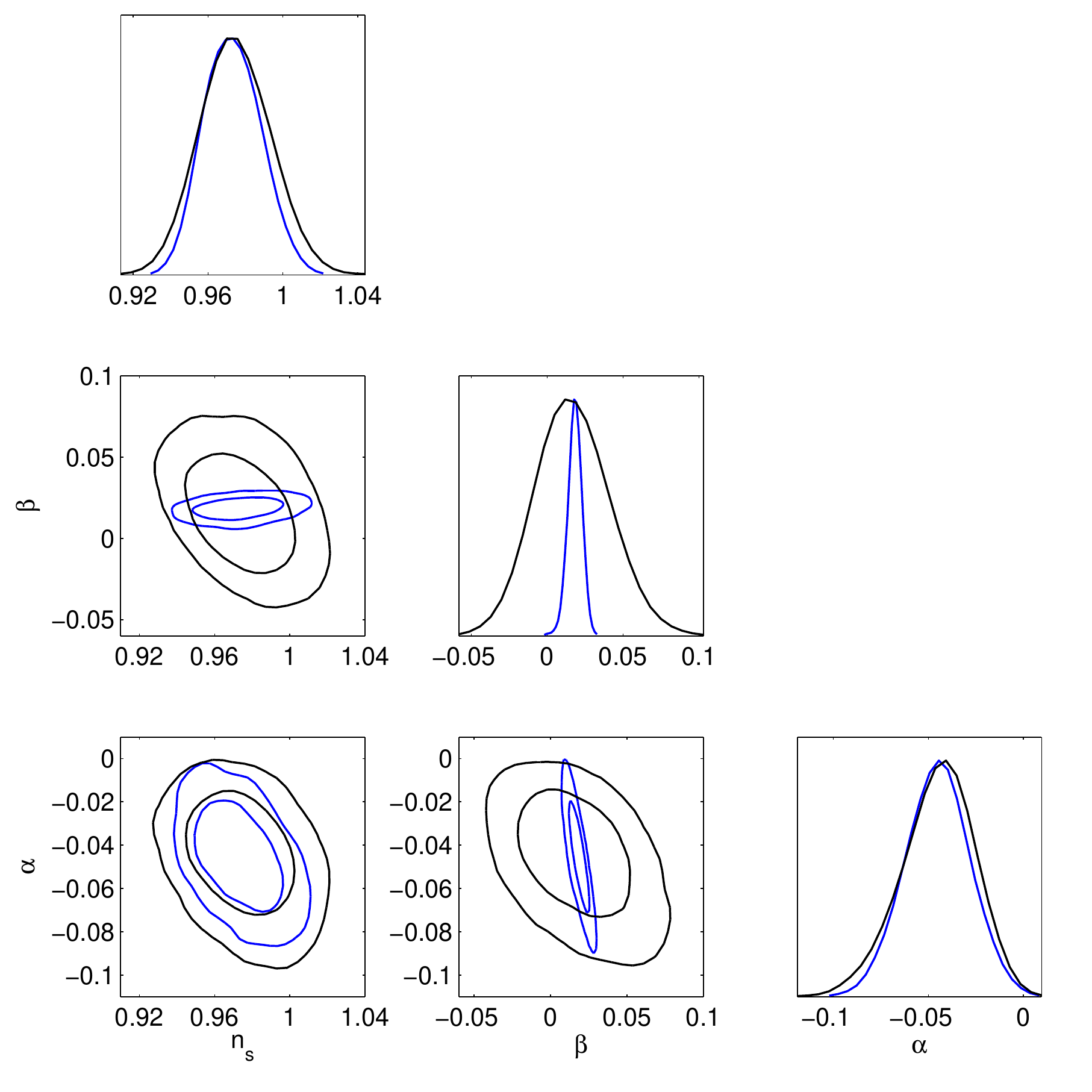}
\caption{Marginalized constraints for $n_s$, $\alpha$, and $\beta$ for WMAP+SPT (black) and
WMAP+SPT+PIXIE (blue) with $\mu_0 = 7.75 \times 10^{-8}$.}
\end{figure*}
Current data is consistent with a null detection of $\beta$, and so our objective is to determine how
large the $\mu$-distortion must be in order to positively detect $\beta$ with PIXIE, {\it i.e.} to substantiate its inclusion as a parameter in the model.  We make
use of two data sources: real cosmic microwave background temperature and polarization anisotropy data,
and a notional $\mu$-distortion signal of the quality expected from PIXIE.  The former data set includes
the latest measurements from the Wilkinson Microwave Anistropy Probe (WMAP) \cite{Komatsu:2010fb} and the South Pole Telescope
(SPT) \cite{Keisler:2011aw}.  These data are well-described by the standard inflationary cosmology with base parameters: the baryon and cold dark matter densities, $\Omega_b h^2$ and $\Omega_c h^2$, the
ratio of the sound horizon to the angular diameter distance at decoupling, $\theta_s$, the
optical depth to reionization, $\tau$, and a power law spectrum of density perturbations parameterized by
amplitude, $A_s$, and spectral tilt, $n_s$.  The SPT data requires three additional foreground parameters:  
the Poisson point source power from randomly distributed galaxies, $D^{\rm PS}_\ell$,
the clustered point source power, $D^{\rm CL}_\ell$, and the secondary emission from the
Sunyaev-Zeldovich effect from clusters, $A_{\rm SZ}$.  The model likelihood calculated from these
parameters, which we denote 
$\mathcal{L}_{C_\ell}$ to indicate that it refers to the WMAP and SPT $C_\ell$ anisotropy spectra, is described in \cite{Komatsu:2010fb,Keisler:2011aw}. 

The second data source, that arising from $\mu$-distortions, is notional.  
For a given fiducial amount of $\mu$-distortion, $\mu_0$, the likelihood of the model with a theoretical
value of $\mu$ given by Eq. (\ref{int}) is computed assuming a Gaussian posterior for $\mu$, 
\begin{eqnarray}
\mathcal{L}_\mu &\propto & p(\mu)/\pi(\mu),\nonumber\\ 
 &\propto& \exp \left[-\frac{(\mu - \mu_0)^2}{2\sigma_\mu^2}\right],
\end{eqnarray}
where the prior, $\pi(\mu)$, is taken to be flat within the range $\mu \in [0,10^{-4}]$,
and $\sigma_\mu = 10^{-9}$ \cite{Pajer:2012vz}.
Then, for a particular set of parameter values, the full likelihood of the model can be determined by simply summing the log-likelihoods of the two data sources,
\begin{equation}
\label{like}
\ln \mathcal{L}_{\rm tot} = \ln \mathcal{L}_{C_\ell} + \ln \mathcal{L}_\mu.
\end{equation} 
To summarize, we are considering an overall likelihood that derives from two separate data sources, CMB anisotropies and $\mu$-distortions, that probe different cosmological scales.  

Since the real CMB anisotropy data are fixed, our task is to vary $\mu_0$ across the prior range until we
achieve $|\ln B_{01}| > 5$.  
This process sets up an interesting tension between the two data sources.  On the one
hand, we have CMB temperature and polarization anisotropy data that is well-fit by the
model $\mathcal{M}_0$ with $\beta = 0$.  In fact, the simpler model with $\alpha = 0$ is
still marginally preferred by current data \cite{Pahud:2007gi}, and is perfectly
consistent with an amount of $\mu$-distortion below the PIXIE detection threshold of
$\mu \sim 10^{-8}$.  However, as the fiducial amount of $\mu$-distortion is increased, 
even the model with running,
$\mathcal{M}_0$, struggles to provide an acceptable fit.  This is illustrated in Figure
1: as $\mu_0$ is increased, $\alpha$ must get larger to accommodate the increase in
small scale power needed to generate the fiducial amount of $\mu$-distortion.  But
$\mathcal{M}_0$ can only provide increased small scale power at the cost of
simultaneously increasing large scale power, and in an effort to fit ever larger values
of $\mu_0$, the model is pulled out of agreement with large scale CMB data.  This effect
signals the need for an additional parameter, $\beta = d\alpha/d\ln k$, that gives a
spectrum with increased small scale power -- to accommodate the large $\mu$-distortion --
without strongly modifying the shape of the large scale power spectrum that is well-constrained by current CMB data. 

In order to calculate $B_{01}$ for a particular value of $\mu_0$, the posterior,
$p(\beta,{\bf d}|\mathcal{M}_1)$, and prior, $\pi(\beta,\mathcal{M}_1)$, distributions must be determined.
The prior encodes our current knowledge of $\beta$, and is taken equivalent to the posterior distribution of $\beta$
obtained from WMAP+SPT data: it is close to Gaussian 
with mean $\bar{\beta} = 0.017$ and standard deviation $\sigma_\beta = 0.024$.  
The posterior distribution is obtained using Markov chain Monte Carlo (MCMC) over
the parameter space of model $\mathcal{M}_1$: $\{\Omega_b h^2,\Omega_c h^2, \theta_s, \tau, A_s, n_s, \alpha, \beta, r, D^{\rm
PS}_\ell, D^{\rm CL}_\ell, A_{\rm SZ}\}$.  
For each sample, Eq. (\ref{int}) is numerically integrated with $P^i(k)$ given by
Eq. (\ref{spec}) to obtain the $\mu$-distortion and the overall
likelihood, $\mathcal{L}_{\rm tot}$, determined from Eq. (\ref{like}).  
This process is iterative: if the $B_{01}$ thus calculated for the
chosen fiducial value of $\mu_0$ does not satisfy $|\ln B_{01}| \approx 5$, a new fiducial $\mu_0$ is
selected and another MCMC run is begun.

In Table 1 we present the results of our analysis for different values of $\mu_0$ in the
range $\mu_0 \in [10^{-8},10^{-7}]$.    
For each value of $\mu_0$, six chains were run using the parameterization of model $\mathcal{M}_1$, and chain convergence was considered achieved when the Gelman-Rubin statistic satisfied $R - 1 < 0.05$.  
\begin{table}[ht] 
\begin{tabular}{c|c}       
\hline                                                                                              
$|\ln B_{01}|$ &$\,\mu_0 \, (\times 10^{-8})$ \\\hline
1&  $6$   \\
\rowcolor{lightgray}                                                                                                     
3.5 & $7.5$ \\
5  & $7.75$\\
\rowcolor{lightgray}                                                                                                     
6.5& 8 \\
\end{tabular}
\caption{Bayes factor, $|\ln B_{01}|$, indicating varying levels of evidence for $\beta \neq 0$ for different fiducial $\mu$-distortion amplitudes,
$\mu_0$.} 
\end{table}
The \texttt{CosmoMC}\footnote{http://cosmologist.info/cosmomc.} software package \cite{Lewis:2002ah} was used for sampling and chain analysis, with
modified likelihood routines to calculate $\mathcal{L}_{\rm tot}$.  We find that
PIXIE will detect $\beta \neq 0$ if $\mu_0 \geq 7.75\times 10^{-8}$, covering much of the PIXIE sensitivity range
($\mu \gtrsim 10^{-8}$).  The corresponding resolution limit of the running of running is relatively large: $\beta
\gtrsim 0.015$.  This is principally a result of the good constraints on $\alpha < 0$ on CMB scales: $\beta$ is forced
to be big to yield a large enough $\mu$-distortion to be detected by an instrument like PIXIE.

In addition to detecting
$\beta \neq 0$, PIXIE will place relatively good constraints on its value: $\beta
= 0.015^{+0.02}_{-0.01}$ at 2$\sigma$, for $\mu_0 = 7.75 \times 10^{-8}$.
Current constraints on the spectral parameters for WMAP+SPT are compared with
projected constraints when PIXIE is included at this level in Figure 2.
It is interesting to
note that while $\mu$-distortion data greatly improves constaints on $\beta$,
there is little change in the posterior distributions of either $n_s$ or $\alpha$
compared to current results.  This is because both $n_s$ and $\alpha$ are already
well-constrained by WMAP+SPT: the existence of sufficient small scale power to
generate a detectable level of $\mu$-distortion establishes the need for $\beta$
and constrains it without affecting the tight limits on $n_s$ and $\alpha$ on CMB
scales. 
\section{Implications for slow roll inflation}
The minimum amount of running of running that can be detected with statistical significance
(strong evidence, or 0.993 posterior odds) is relatively large: $\beta \gtrsim 0.015$, possibly
in conflict with slow roll inflation, provided that the observed $\mu$-distortion is primordial.  The slow roll approximation holds whenever the acceleration
of the scalar field, $\ddot{\phi}$, can be neglected relative to the cosmological drag,
$\ddot{\phi} \ll 3H\dot{\phi}$, with the result that
\begin{equation}
3H\dot{\phi} \approx -V',
\end{equation}
where dots denote time derivatives and primes denote derivatives with respect to $\phi$.  This is
the case as long as the scalar field is potential-dominated: $\dot{\phi}^2/2 \ll V(\phi)$. Slow
roll can be maintained as long as the potential is sufficiently flat; this condition is often
stated in terms of the smallness of the parameters,  
\begin{eqnarray}
\epsilon &\equiv & 2M_{\rm Pl}^2 \left(\frac{H'}{H}\right)^2 \approx \frac{M^2_{\rm Pl}}{2}\left(\frac{V'}{V}\right)^2 \ll 1 \nonumber \\
\eta &\equiv& 2M_{\rm Pl}^2 \frac{H''}{H} \approx \frac{M_{\rm Pl}^2}{2}\frac{V''}{V}\ll
\mathcal{O}(1)\nonumber \\
&&\vdots \nonumber
\end{eqnarray}
where the dots denote terms involving higher derivatives of $H(\phi)$.  But how many terms must we
consider?  How small do they need to be?
The slow roll approximation can be formally derived by considering the Taylor expansion of the Hubble
parameter \cite{Liddle:1994dx},
\begin{equation}
\label{H}
\frac{H(\phi)}{H_0} = 1 +
\sum_{\ell =0}  \frac{{}^\ell \lambda_{H0}}{2^\ell (\ell+1)!B_1^{\ell-1}}\left(\frac{\phi}{M_{\rm Pl}}\right)^{\ell+1},
\end{equation}
where $B_1 = \sqrt{\epsilon_0/2}$ and the subscript ``0'' indicates that the coefficients should be
evaluated when the pivot scale, $k_0$, exits the horizon, and we have chosen $\phi_0 = 0$.  The coefficients,
\begin{equation}
\label{flow}
{}^\ell \lambda_H \equiv (2M_{\rm Pl})^\ell
\frac{(H')^{\ell-1}}{H^\ell}\frac{d^{(\ell+1)}H}{d\phi^{(\ell+1)}},\,\,\,\ell \geq 0
\end{equation}
generate the hierarchy of slow roll parameters derived from $H(\phi)$: ${}^1 \lambda_H = \eta$,
${}^2 \lambda_H = \xi^2$, etc.  
In this representation, a cosmological solution given by Eq. (\ref{H}) truncated at some order $M$ is interpreted as a perturbation expansion
about de Sitter space. 

The parameters $\epsilon$ and the ${}^\ell \lambda_H$ determine approximate expressions for the spectral
parameters: $n_s$, $\alpha$, $\beta$, $\cdots$, that are themselves coefficients of the Taylor
expansion of $\ln P(k)$ in $\ln k$.   The lowest order spectral parameter, $n_s$, is lowest order in
slow roll, $n_s(\epsilon, \eta, \cdots)$, where $\cdots$ includes all higher-order terms $\epsilon^2,
\epsilon \eta, \eta^2$, and so on.  The second order spectral parameter, $\alpha$, is likewise
second order in slow roll, $\alpha(\xi^2, \epsilon^2, \epsilon \eta,\cdots)$.  The slow
roll approximation can be considered valid if the n$^{th}$-order spectral parameter is dominated by the
n$^{th}$-order slow roll parameters, with higher-order terms comprising ever smaller corrections.
It is therefore a prediction of slow roll inflation that higher-order spectral parameters should be
vanishingly small, and that a general solution should be described by only the few lowest-order
terms.

We begin by writing the spectral observables as follows \cite{Stewart:1993bc,Gong:2001he,Huang:2006yt},
\begin{eqnarray}
n_s &=& 1-4\epsilon + 2\eta -2C\xi^2 -8(C+1)\epsilon^2  +(6+10C)\epsilon\eta \nonumber \\ 
\alpha &=& -2\xi^2 -8\epsilon^2+10\epsilon \eta - (8+14C)\epsilon \xi^2 \nonumber \\ &&+ 2C\eta \xi^2 +
2C\,{}^3\lambda_H \nonumber \\
\beta &=& -14\epsilon \xi^2 + 2\eta \xi^2 -2C\xi^4 + 2\,\,{}^3\lambda_H - 2C\,\,{}^4\lambda_H \nonumber \\ && -
32\epsilon^3 + 62\epsilon^2 \eta -20\epsilon \eta^2\nonumber - (46+56C)\epsilon^2\xi^2 \nonumber \\ &&+
(26+48C)\epsilon\eta \xi^2\nonumber  -2C\eta^2\xi^2 + (10+20C)\epsilon\,\, {}^3\lambda_H\nonumber \\
&& - 6C\eta
\,\,{}^3\lambda_H,
\end{eqnarray}
where $C \simeq -0.73$.
The subscript ``0'' has been omitted for brevity; it should be understood that
the above expressions are to be evaluated at the pivot scale, $k_0$, and associated field
value, $\phi_0$.  
We have written the observables to next-to-leading order in the slow roll parameters because 
we would like to study the contributions of these higher-order terms, especially for $\beta$.  
We first set ${}^4\lambda_H = 0$, truncating the hierarchy at order $M=3$.  We then seek a
cosmological solution, parameterized by
$\{\epsilon(\phi),\eta(\phi),\xi^2(\phi),\,{}^3\lambda_H(\phi)\}$ that yields observables in
agreement with current data in combination with a discovery of $\mu$-distortion by PIXIE with
$\mu_0 = 7.75\times 10^{-8}$ (cf. Figure 2.)  

\begin{figure*}[ht]
$\begin{array}{cc}
\subfigure[]{
\includegraphics[width=0.5 \textwidth]{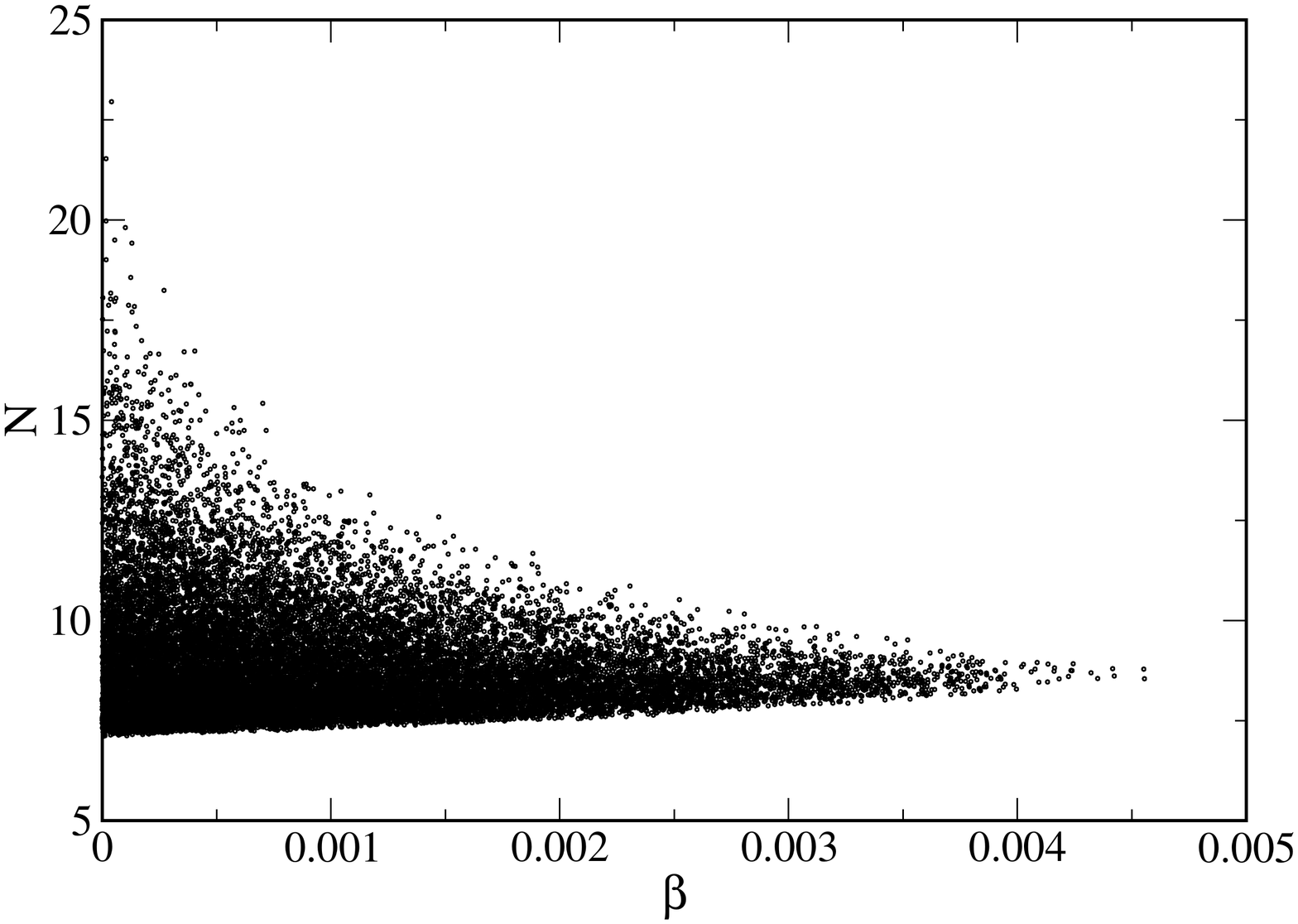}}
\subfigure[]{
\includegraphics[width=0.5 \textwidth]{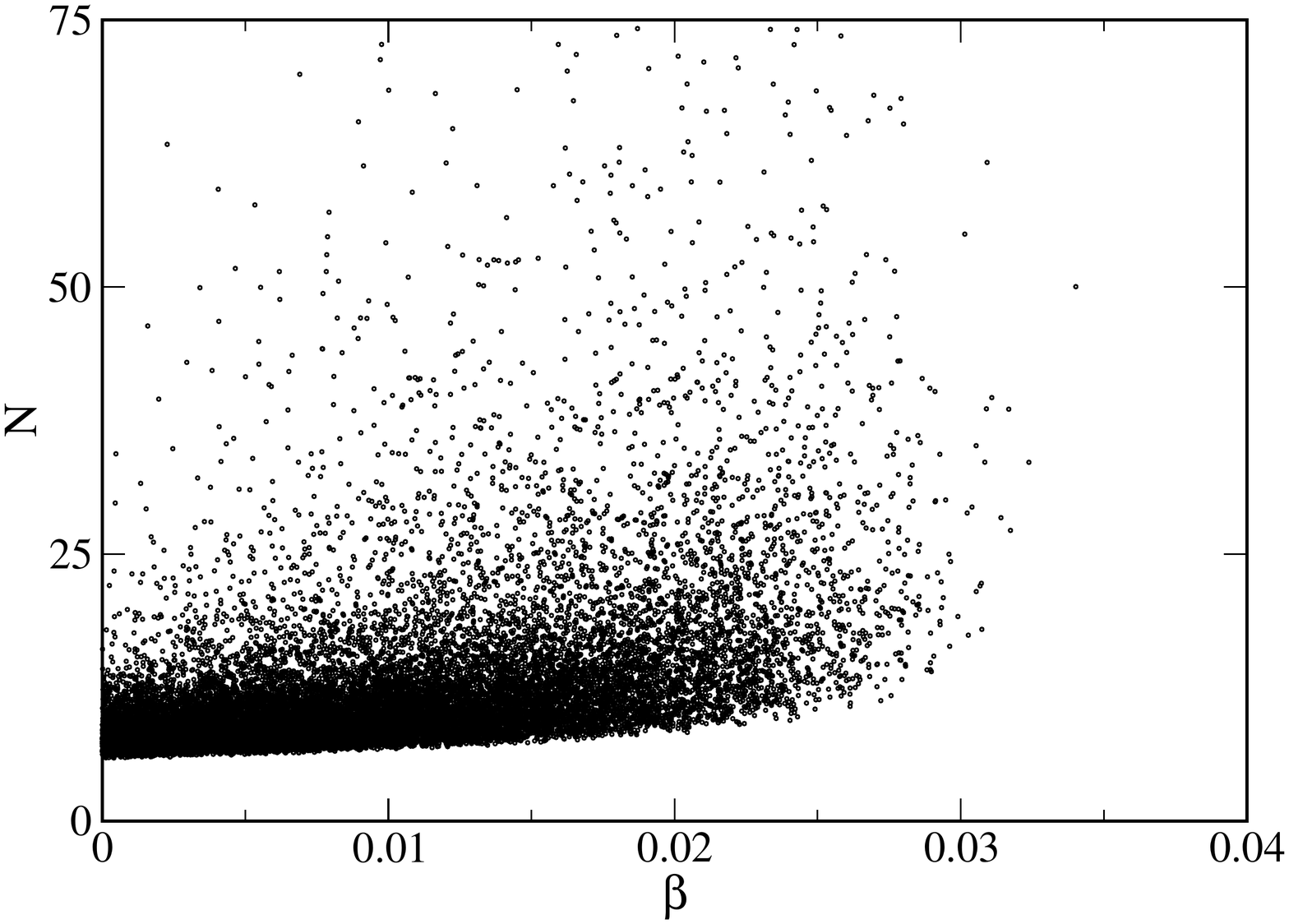}}
\end{array}$
\caption{(a) Distribution of models generated stochastically at third order in slow roll.  The models are in
agreement with current CMB data, but they fail to provide a sufficient number of e-folds, $N$, of inflation or yield large enough $\beta$. (b)
Distribution of models generated stochastically at fourth order in slow roll.  While most models still provide
insufficient inflation, many are successful with large $\beta$.}
\end{figure*}
In addition to satisfying parameter constraints, the solution must yield sufficient inflation
to solve the horizon and flatness problems: $N \gtrsim 30$ e-folds if inflation is to occur
above the electroweak scale, and $N \approx 55$ $(60)$ for GUT (SUSY) scale inflation.  The
number of e-folds of expansion is given by 
\begin{equation}
\label{N}
\frac{dN}{d\phi} = \frac{1}{2M_{\rm Pl}^2}\frac{H}{H'},
\end{equation}
where $H(\phi)$ is determined by the slow roll parameters up to order $M=3$ via Eq.
(\ref{H}).  For a given set of parameters $\{\epsilon_0,\eta_0,\xi^2_0,\,{}^3\lambda_{H0}\}$
Eq. (\ref{N}) can be integrated numerically to obtain the number of e-folds of expansion that
occur after the scale $k_0$ leaves the horizon\footnote{This is equivalent to solving the
system of flow equations truncated at order $M=3$ \cite{Hoffman:2000ue,Kinney:2002qn}.}.  We
stochastically sampled the $\epsilon-\eta-\xi^2-\,{}^3\lambda_H$ space, and integrated Eq. (\ref{N}) for each set of initial
conditions.  We collected
solutions with $r < 0.3$, $n_s \in [0.94,1.01]$, and $\alpha \in [-0.08,0]$, ranges that approximate the 95\%
confidence limits obtained on the parameters in the previous section.  No constraints were placed on $N$ or $\beta$
so that the full space could be explored.  As we show
in Figure 3 (a), not only do the models fail to provide enough inflation\footnote{A similar investigation was
carried out in
\cite{Easther:2006tv} regarding the implications for slow roll inflation of $\alpha$ near the WMAP3 centroid,
$\alpha \approx -0.05$.}, they all give $\beta
\lesssim 0.005$ -- just outside the  2$\sigma$ limit.  This result shows that if $\beta$ is
controlled only by leading-order terms in the slow roll expansion, it is not possible to find a working
single field inflation model that fits observations.

What about the next-to-leading order term, ${}^4\lambda_H$?  If we truncate the slow roll hierarchy at
order $M=4$, our analysis shows that not only is sufficient inflation obtained, but $\beta$ can now be large
(Figure 3 (b)).  One such model is given by the parameter values: $\{\epsilon_0, \eta_0, \xi^2_0, \,\,{}^3\lambda_H, \,\,{}^4\lambda_H\} = \{0.015,-0.01,0.035,0.018,0.0025\}$.  Not only are the first four parameters of equal magnitude, but the fifth parameter,
${}^4\lambda_H$,
while only amounting to a small correction according to the slow roll
approximation, strongly affects the inflationary dynamics and
allows for ${}^3\lambda_H$ to become large.  Our analysis therefore shows that for $\beta \gtrsim 0.005$, the slow roll expansion must be taken to at least
fourth order to obtain viable solutions, many of which are described by slow roll parameters of roughly equal magnitude and
relevance. Given this observation, there is little reason to expect that
higher-order terms can be safely neglected. 
We conclude that a statistically significant detection of running of
running by PIXIE, $\beta \gtrsim 0.015$, is in
conflict with single field, slow roll inflation.

\section{Conclusions} 
The discovery of higher-order runnings of the scalar power spectrum has important
implications for the nature of the inflationary era.  Large higher-order runnings threaten
the validity of the slow
roll approximation, and disfavor an explanation in terms of single field, slow roll
inflation. Current data from WMAP+SPT have almost conclusively detected $\alpha \neq 0$, and Planck will furnish a
detection if the value of $\alpha$ is close to current best-fit estimate, $\alpha \approx -0.024$.  Meanwhile, the
running of running, $\beta = d\ln \alpha/d \ln k$, is not well constrained because today's observatories do not
probe the primordial spectrum, $P(k)$, across a sufficiently large range of length scales to resolve it.      
In this work, we investigated whether a future measurement of CMB $\mu$-distortions by an experiment like PIXIE,
taken in combination with CMB anisotropy data, would provide a sufficiently long
lever arm to resolve these higher-order runnings.

We framed the detectability of $\beta$ as a problem in Bayesian model comparison between two power spectra: one with
$\beta$ and one without.  We computed the Bayes factor using MCMC for different fiducial values of $\mu_0$ in order
to determine how large the notional $\mu$-distortion signal needs to be to favor the running of running model.
We find that if $\mu_0 > 7.75 \times 10^{-8}$, the inclusion of $\beta$ in the model is required by the data with
strong significance (equivalent to 0.993 posterior odds.), covering much of the PIXIE sensitivity range.  The
best-fit value associated with detection is $\beta \approx 0.015$, which is relatively large given any presumption
of slow roll.  This is a result of the tight constraints on $\alpha$ on CMB scales: with $\alpha < 0$, $\beta$ must
be moderately large and positive in order for the spectrum to produce enough $\mu$-distortion on small scales to be
detectable by PIXIE.  Constraints on $\beta$
are also improved by around a factor of 5 over current estimates.  

We next studied the implications that large $\beta$ has for single field slow roll
inflation under the assumption that the primordial distortion is distinct from post-inflationary
sources.  We performed a flow analysis, focusing particularly on the next-to-leading order terms in the
slow roll expansion and their role in the inflationary dynamics and observables.
It was found that  models taken to leading
order ($M=3$) failed to yield sufficient inflation or satisfy constraints on $\beta$, while a truncation at
next-to-leading order ($M=4$) gave models with very different inflationary trajectories that succeeded in meeting these criteria.  The inclusion of ${}^4
\lambda_H$, while amounting to only a small correction according to the
slow roll approximation, strongly alters
the inflationary dynamics.  Successful $M=4$ models had slow roll parameters of roughly equal magnitude and
importance, challenging the slow roll prescription that higher-order terms in the expansion should indicate ever
smaller corrections to the inflationary observables and trajectories.  We conclude that a detection of running of running by PIXIE
will be in conflict with single field, slow roll inflation.  

The search for CMB $\mu$-distortions is a promising avenue for challenging
the prevailing conception of inflation.  It is important that work
continues on understanding the possible foregrounds that might contaminate
any $\mu$-distortion signal, as well as identifying possible
post-inflationary sources of $\mu$-distortions that might mimic a
primordial signal.  If a measurement of primordial $\mu$-distortions by
PIXIE 
results in a detection of $\beta$, then this will be a strong
indication that inflation either exhibits non-slow roll dynamics or is
driven by multiple fields. 
\acknowledgments
The author would like to thank Will Kinney, Jens Chluba, and Enrico Pajer for comments, and Simon DeDeo and Damien Easson for helpful discussions.


\begin{thebibliography}{999}
\bibitem{Vazquez:2012ux} 
  J.~A.~Vazquez, M.~Bridges, M.~P.~Hobson and A.~N.~Lasenby,
  JCAP {\bf 1206}, 006 (2012)
  [arXiv:1203.1252 [astro-ph.CO]].
\bibitem{Lesgourgues:2007gp} 
  J.~Lesgourgues and W.~Valkenburg,
  Phys.\ Rev.\ D {\bf 75}, 123519 (2007)
  [astro-ph/0703625].
\bibitem{Powell:2007gu} 
  B.~A.~Powell and W.~H.~Kinney,
  JCAP {\bf 0708}, 006 (2007)
  [arXiv:0706.1982 [astro-ph]]
\bibitem{Lidsey:1995np} 
  J.~E.~Lidsey, A.~R.~Liddle, E.~W.~Kolb, E.~J.~Copeland, T.~Barreiro and M.~Abney,
  Rev.\ Mod.\ Phys.\  {\bf 69}, 373 (1997)
  [astro-ph/9508078].
\bibitem{Komatsu:2010fb} 
  E.~Komatsu {\it et al.}  [WMAP Collaboration],
  Astrophys.\ J.\ Suppl.\  {\bf 192}, 18 (2011)
  [arXiv:1001.4538 [astro-ph.CO]].
\bibitem{Viel:2004np} 
  M.~Viel, J.~Weller and M.~Haehnelt,
  Mon.\ Not.\ Roy.\ Astron.\ Soc.\  {\bf 355}, L23 (2004)
  [astro-ph/0407294].
\bibitem{McDonald:2004xn} 
  P.~McDonald {\it et al.}  [SDSS Collaboration],
  Astrophys.\ J.\  {\bf 635}, 761 (2005)
  [astro-ph/0407377].
\bibitem{Seljak:2006bg} 
  U.~Seljak, A.~Slosar and P.~McDonald,
  JCAP {\bf 0610}, 014 (2006)
  [astro-ph/0604335].
\bibitem{Dunkley:2010ge}
  J.~Dunkley {\it et al.},
  arXiv:1009.0866 [astro-ph.CO]
\bibitem{Keisler:2011aw} 
  R.~Keisler, C.~L.~Reichardt, K.~A.~Aird, B.~A.~Benson, L.~E.~Bleem, J.~E.~Carlstrom, C.~L.~Chang and
H.~M.~Cho {\it et al.},
  Astrophys.\ J.\  {\bf 743}, 28 (2011)
  [arXiv:1105.3182 [astro-ph.CO]].
\bibitem{Loeb:2003ya} 
  A.~Loeb and M.~Zaldarriaga,
  Phys.\ Rev.\ Lett.\  {\bf 92}, 211301 (2004)
  [astro-ph/0312134].
\bibitem{Barger:2008ii} 
  V.~Barger, Y.~Gao, Y.~Mao and D.~Marfatia,
  Phys.\ Lett.\ B {\bf 673}, 173 (2009)
  [arXiv:0810.3337 [astro-ph]].
\bibitem{Adshead:2010mc} 
  P.~Adshead, R.~Easther, J.~Pritchard and A.~Loeb,
  JCAP {\bf 1102}, 021 (2011)
  [arXiv:1007.3748 [astro-ph.CO]].
\bibitem{Kogut:2011xw} 
  A.~Kogut, D.~J.~Fixsen, D.~T.~Chuss, J.~Dotson, E.~Dwek, M.~Halpern, G.~F.~Hinshaw and S.~M.~Meyer
{\it et al.},
  JCAP {\bf 1107}, 025 (2011)
  [arXiv:1105.2044 [astro-ph.CO]].
\bibitem{Khatri:2011aj} 
  R.~Khatri, R.~A.~Sunyaev and J.~Chluba,
  Astron.\ Astrophys.\  {\bf 540}, A124 (2012)
  [arXiv:1110.0475 [astro-ph.CO]].
\bibitem{Chluba:2012gq} 
  J.~Chluba, R.~Khatri and R.~A.~Sunyaev,
  arXiv:1202.0057 [astro-ph.CO].
\bibitem{Dent:2012ne} 
  J.~B.~Dent, D.~A.~Easson and H.~Tashiro,
  arXiv:1202.6066 [astro-ph.CO].
\bibitem{Chluba:2012we} 
  J.~Chluba, A.~L.~Erickcek and I.~Ben-Dayan,
  arXiv:1203.2681 [astro-ph.CO].
\bibitem{Silk:1967kq} 
  J.~Silk,
  Astrophys.\ J.\  {\bf 151}, 459 (1968).
\bibitem{Sunyaev:1970er} 
  R.~A.~Sunyaev and Y.~.B.~Zeldovich,
  Astrophys.\ Space Sci.\  {\bf 7}, 20 (1970).
\bibitem{Peebles:1970ag} 
  P.~J.~E.~Peebles and J.~T.~Yu,
  Astrophys.\ J.\  {\bf 162}, 815 (1970).
\bibitem{Illarionov:1975}
  A. F. Illarionov and R. A. Sunyaev,
  Soviet Astr., 18, 413 (1975)
\bibitem{Coles}
  P. Coles and J. Barrow,
  Mon. Not. R. astr. Soc {\bf 244}, 188 (1990)
\bibitem{Daly}
  R. A. Daly,
 Astrophys. J. {\bf 371}, 14 (1991)
\bibitem{Hu:1992dc} 
  W.~Hu and J.~Silk,
  Phys.\ Rev.\ D {\bf 48}, 485 (1993).
\bibitem{Chluba:2011hw} 
  J.~Chluba and R.~A.~Sunyaev,
  arXiv:1109.6552 [astro-ph.CO].
\bibitem{Danese:1982}
L. Danese and G. De Zotti
Astron. Astrophys. 107, 39-42 (1982)
\bibitem{Jeffreys}
H. Jeffreys, {\it Theory of Probability}, 3rd ed, Oxford University Press (1961)
\bibitem{Trotta:2005ar} 
  R.~Trotta,
  Mon.\ Not.\ Roy.\ Astron.\ Soc.\  {\bf 378}, 72 (2007)
  [astro-ph/0504022].

\bibitem{Dickey:1971}
  J. M.~Dickey,
  Ann. Math. Stat., 42, 204 (1971)
\bibitem{Pajer:2012vz} 
  E.~Pajer and M.~Zaldarriaga,
  arXiv:1201.5375 [astro-ph.CO].
\bibitem{Pahud:2007gi} 
  C.~Pahud, A.~R.~Liddle, P.~Mukherjee and D.~Parkinson,
  Mon.\ Not.\ Roy.\ Astron.\ Soc.\  {\bf 381}, 489 (2007)
  [astro-ph/0701481].
\bibitem{Lewis:2002ah} 
  A.~Lewis and S.~Bridle,
  Phys.\ Rev.\ D {\bf 66}, 103511 (2002)
  [astro-ph/0205436].
\bibitem{Liddle:1994dx}
  A.~R.~Liddle, P.~Parsons and J.~D.~Barrow,
  Phys.\ Rev.\  D {\bf 50}, 7222 (1994)
  [arXiv:astro-ph/9408015].
\bibitem{Stewart:1993bc} 
  E.~D.~Stewart and D.~H.~Lyth,
  Phys.\ Lett.\ B {\bf 302}, 171 (1993)
  [gr-qc/9302019].
\bibitem{Gong:2001he} 
  J.~-O.~Gong and E.~D.~Stewart,
  Phys.\ Lett.\ B {\bf 510}, 1 (2001)
  [astro-ph/0101225].
\bibitem{Huang:2006yt} 
  Q.~-G.~Huang,
  Phys.\ Rev.\ D {\bf 76}, 043505 (2007)
  [astro-ph/0610924].
\bibitem{Hoffman:2000ue} 
  M.~B.~Hoffman and M.~S.~Turner,
  Phys.\ Rev.\ D {\bf 64}, 023506 (2001)
  [astro-ph/0006321].
\bibitem{Kinney:2002qn} 
  W.~H.~Kinney,
  Phys.\ Rev.\ D {\bf 66}, 083508 (2002)
  [astro-ph/0206032].
\bibitem{Easther:2006tv} 
  R.~Easther and H.~Peiris,
  JCAP {\bf 0609}, 010 (2006)
  [astro-ph/0604214].
\end{thebibliography}
\end{document}